\documentclass[aps,prl,superscriptaddress,floatfix,nofootinbib, twocolumn]{revtex4}
\usepackage{mathrsfs}
\usepackage{epsfig,psfrag}
\usepackage{amsmath,amsfonts,amssymb, latexsym}
\usepackage{epsfig}
\usepackage[usenames]{color}
\usepackage{subfigure}
\usepackage{bm}

\definecolor{BrickRed}{cmyk}{0,0.89,0.94,0.28}
\definecolor{MidnightBlue}{cmyk}{0.98,0.13,0,0.43}
\definecolor{DarkGreen}{rgb}{0,0.7,0.1}

\definecolor{RedViolet}{cmyk}{0.07,0.90,0,0.34}
\definecolor{SeaGreen}{cmyk}{0.69,0,0.50,0}
\definecolor{FireOrange}{rgb}{1.,.294,.247}

\begin{document}
\preprint{draft}
\title{Casimir interactions of an object inside a spherical metal shell}
\author{Saad Zaheer\footnote{Present address: Department of Physics and Astronomy, University of Pennsylvania, Philadelphia, PA 19104-6396, USA}}
\affiliation{Department of Physics, Massachusetts Institute of Technology, Cambridge, MA 02139, USA}
\author{Sahand Jamal Rahi}
\affiliation{Department of Physics, Massachusetts Institute of Technology, Cambridge, MA 02139, USA}
\author{Thorsten Emig}
\affiliation{Institut f\"ur Theoretische Physik, Universit\"at zu K\"oln, Z\"ulpicher Strasse 77, 50937 K\"oln, Germany}
\affiliation{Laboratoire de Physique Th\'eorique et Mod\`eles Statistiques, CNRS UMR 8626, Universit\'e Paris-Sud, 91405 Orsay, France}
\author{Robert L. Jaffe}
\affiliation{Center for Theoretical Physics, Laboratory for Nuclear Science, and Department of Physics, Massachusetts Institute of Technology, Cambridge, MA 02139, USA}
\begin{abstract}
  We investigate the electromagnetic Casimir interactions of an object
  contained within an otherwise empty, perfectly conducting
  spherical shell.  For a small object we present analytical
  calculations of the force, which is directed away from the center of
  the cavity, and the torque, which tends to align the object opposite
  to the preferred alignment outside the cavity.  For a perfectly conducting sphere as the interior
  object, we compute the corrections to  the proximity force approximation (PFA) numerically.  In both cases  the results for the interior configuration match smoothly onto those  for the corresponding exterior configuration.
\end{abstract}
\maketitle

The development of methods to measure Casimir forces with high precision~\cite{KITP}
and the advent of microelectromechanical devices, for which Casimir
forces are important~\cite{Chan01,Capasso07}, have stimulated the
development of new theoretical tools for analyzing these forces for
geometries beyond parallel plates~\cite{casimir}. In particular,
methods based on scattering theory~\cite{thorsten,scalar,scattering}
have been used to precisely compute Casimir forces and torques for a
wide variety of configurations of both perfect conductors and
dielectrics.  We have applied these methods to find the electrodynamic
Casimir interaction of a conducting or dielectric object inside a
perfectly conducting spherical cavity~\cite{saad}. Earlier studies of
Casimir interactions of one object contained within another have been
limited to infinitely long cylinders~\cite{cylinders,Rodriguez08}.

In this paper we report two sets of results. First, we determine the
forces and torques on a small object, dielectric or conducting, well
separated from the cavity walls. This is the interior analogue of the
famous Casimir-Polder force on a polarizable molecule near a perfectly
conducting plate~\cite{casimirpolder}.  Second, we compute the
interaction energy of a finite-size metal sphere with the cavity walls
when the separation, $d$, between their surfaces tends to zero. The
leading term in $d$ is known to be given by the PFA~\cite{PFA}. By
evaluating our closed-form expressions numerically, we find the next
to leading term in an expansion in $d$. The main achievements of our
analysis are: (i) The first exact results for the
electromagnetic interaction of an interior macroscopic object
with a confining cavity; (ii) The determination of the
  corrections to the PFA for two spheres for all values of their
  radii, both positive (exterior) and negative (interior); (iii) A
physical understanding of the absence of an orientation dependence of
the Casimir-Polder (CP) force between an object and a plane mirror;
and (iv) The generalization of the CP force to concave
mirrors.
The  last follows from our analysis
when the cavity radius is much larger than $d$ and the size of the
interior object. Then the interaction is determined only by a concave
segment of the shell that is closest to the object and our results are
expected to apply also to this open geometry. Our work suggests that 
the  orientation dependence of Casimir forces near curved surfaces might
prove to be an interesting area for future experiments.  Furthermore,
the interactions described here are relevant for trapped atoms in
spherical optical resonators \cite{An94}.

An understanding of the corrections to the PFA has been sought for
some time~\cite{beyondpfa}. 
Our works extends beyond previous attempts, many of which have treated
hypothetical scalar fields, by considering electromagnetic fields for  
two objects with different curvature and relative position.
For the sphere-plane
geometry~\cite{thorstencyl,Maia_Neto08} and two spheres of
\emph{equal} radii facing each other~\cite{thorsten} the corrections
have recently been computed.  We repeat these computations for other
ratios of radii of two spheres outside one another and find that the
inside and outside results connect smoothly. This is of direct
experimental significance because Casimir force measurements are
usually performed with spherical rather than perfectly planar
surfaces.

Our analysis is facilitated by a recently developed formalism that combines path integral and scattering theory techniques and is applicable to general shapes and configurations, including objects inside each other~\cite{thorsten, scalar, jamalem}. In the case where an object, $\Sigma_{i}$, lies inside a perfectly conducting cavity with inner surface $\Sigma_{o}$, the Casimir energy is given by~\cite{saad,jamalem}, 
\begin{align}
\mathcal{E} &= \frac{\hbar c}{2 \pi } \int_0^{\infty} d\kappa \ln \frac{\det (\mathcal{I} -  \mathcal{F}^{-1}_{o}  \mathcal{W}^{io}  \mathcal{F}_{i}  \mathcal{V}^{io})}{\det (\mathcal{I}- \mathcal{F}^{-1}_o  \mathcal{F}_i)}.
\label{eq:master}
\end{align}
$\mathcal{F}_{i}$ is the scattering amplitude for electromagnetic waves off the interior object; $\mathcal{F}^{-1}_{o}$ is the inverse scattering amplitude for the conducting cavity, a sphere in our case.  The appearance of the inverse of $\mathcal{F}_{o}$ is a result of the interior geometry~\cite{jamalem}. These scattering amplitudes are matrices evaluated in a spherical vector wave basis with respect to appropriately chosen origins within each object.  The translation matrices, $\mathcal{W}^{io}$ and $\mathcal{V}^{io}$, relate regular wave functions between the coordinate systems of the interior object and the spherical cavity. ($\mathcal{W}^{io} \sim \mathcal{V}^{io \dagger}$ up to multiplication by $(-1)$ of some matrix elements, see Ref.~\cite{jamalem} for details.) All of these matrices are functions of the imaginary frequency $\omega=i\kappa$. The determinant in the denominator subtracts the Casimir energy when the origins of the two objects coincide. This way of normalizing the Casimir energy differs from the exterior case where the objects are removed to infinite separation; a choice that would be unnatural in the interior case. 

Let us first consider the limit, in which the interior object is much smaller than the radius of the spherical cavity. Using the matrix identity $\ln \det \mathcal{M} = \text{Tr} \ln\mathcal{M}$, the integrand in Eq.~(\ref{eq:master}) can be expanded as, $\mathcal{E} = -\hbar c/(2\pi)\int_0^\infty d\kappa \,  \rm{Tr}( \mathcal{N} + \mathcal{N}^2/2 +\ldots)$, where $\mathcal{N} = \mathcal{F}^{-1}_{o}  \mathcal{W}^{io}  \mathcal{F}_{i}  \mathcal{V}^{io}$ describes a reflection event where a wave travels from one object to the other and back~\cite{thorsten}. In general, all terms in the series expansion are important, illustrating the fundamentally non-two-body nature of the Casimir force. The rate of convergence of this series depends on the size of $\Sigma_i$ relative to the separation of its surface from that of $\Sigma_o$.  

If the inner object is small compared to the size of the cavity, the first term in the series, $\text{Tr }\mathcal{N}$, already gives an excellent approximation to the energy. In the small-size limit, the scattering amplitude $\mathcal{F}_{i,lmP,l'm'P'}$, (where $l$ and $m$ are angular momentum indices and $P$ labels $M$ or $E$ polarization) can be expanded in powers of $\kappa$. Only the following terms contribute to lowest order: $\mathcal{F}_{i,1mP,1m'P}(\kappa) = 2\kappa^{3}\alpha^{P}_{mm'}/3 + O(\kappa^4)$, where $\alpha^{P}_{mm'}$ is the static electric ($P=E$) or magnetic ($P=M$) polarizability of the inner object, the same tensor that determines the Casimir-Polder interaction~\cite{casimirpolder,Feinberg68,orientation}.

\begin{figure}[t] 
\includegraphics[width= 8.6cm]{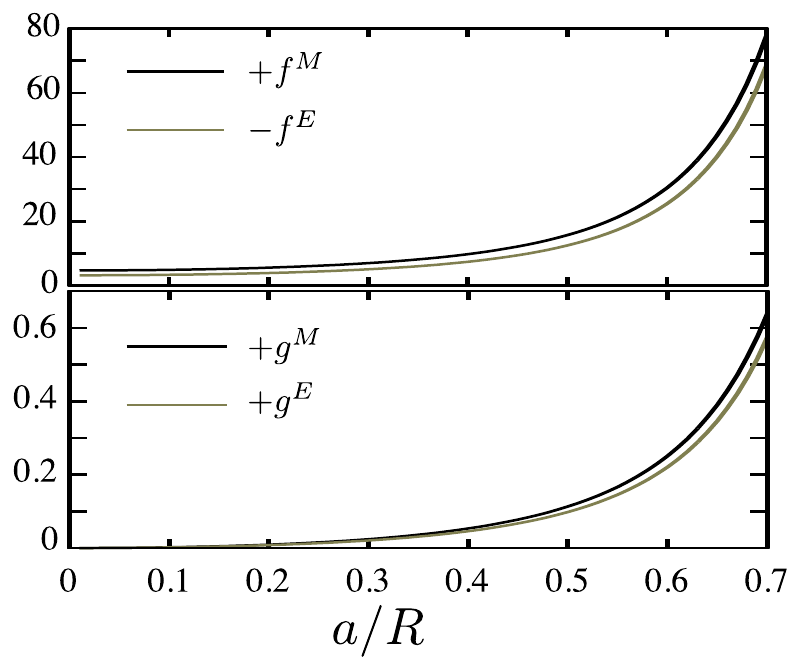}\caption{\label{dfunction}Plot of the functions $f^{M/E}(a/R)$ and $g^{M/E}(a/R)$, defined in Eqs.~(\ref{eq:trace1}) and (\ref{eq:trace2}), respectively.}
\label{hdme}
\end{figure}

We fix $\Sigma_o$ to be a spherical shell of radius $R$ and define $a$ to be the displacement of the center of the interior object from the center of the sphere. We find for the Casimir energy to leading order in $r/R$ (where $r$ is the typical length scale of the interior object), the interior analog of the Casimir-Polder interaction~\cite{saad}, 
\begin{align}
  &\frac{3\pi R^{4}}{\hbar c} \mathcal{E}(a/R)     = \left[f^{E}(a /R)-f^E(0)\right]  {\text{Tr} \alpha^E} \nonumber \\  & + g^{E}(a/R) (2 \alpha^E_{zz}-\alpha^E_{xx}- \alpha^E_{yy})   + (E \leftrightarrow M). \label{eq:orientation}
\end{align}
The $z$-axis is oriented from the center of $\Sigma_o$ to $\Sigma_i$,
and $\alpha^{P}_{ij}$ represent the interior object's static
polarizability tensors in a Cartesian basis. Corrections to this
energy come from dynamic (frequency dependent) dipole polarizabilities
and higher order multipoles \cite{Au+72}. The coefficient
functions $f^P$ and $g^P$, plotted in Fig.~\ref{hdme}, can be
expressed in terms of modified Bessel functions $I_\nu$ and $K_\nu$ as
(with $\lambda = l-\frac{1}{2}, \mu = l+\frac{3}{2}$ for brevity),
\begin{align}
f^E(y) &= \int_0^\infty dx x^3 \sum_{l=1}^{\infty} \Bigg[ \frac{\zeta_l^E(x)}{2xy} \left((l+1)I^2_{\lambda}(xy) + lI^2_{\mu}(xy)\right) \nonumber \\ &-\zeta_l^M(x) \frac{ xy}{2(2l+1)}\left(I_{\lambda}(xy) - I_{\mu}(xy)\right)^2\Bigg] \label{eq:trace1}\\
g^E(y) &= \int_0^\infty dx x^3 \sum_{l=1}^{\infty} \Bigg[\frac{\zeta_l^E(x)}{2xy(2l+1)}\Big(\frac{l^2-1}{2}I^2_{\lambda}(xy) \nonumber \\ &+ \frac{l(l+2)}{2}I^2_{\mu}(xy) - 3l(l+1)I_{\lambda}(xy)I_{\mu}(xy) \Big) \nonumber  \\
&+ \zeta_l^M(x)\frac{xy}{4(2l+1)} \left(I_{\lambda}(xy) - I_{\mu}(xy)\right)^2\Bigg] \, ,\label{eq:trace2}
\end{align}
and $f^M$ and $g^M$ are obtained by substituting $(E \leftrightarrow M)$ in the above equations. The functions $\zeta_l^{M/E}$ are given by (with $\eta = l+1/2$)
\begin{align}
\zeta_l^M(x) = \frac{K_{\eta}(x)}{I_{\eta}(x)}, 
\zeta_l^E(x) = \frac{K_{\eta}(x) + 2xK'_{\eta}(x)}{I_{\eta}(x) + 2xI'_{\eta}(x)} \, .\label{eq:gamma}
\end{align}

$f^E$ is negative and decreasing with $a/R$, while $f^M$ is positive and increasing. Notice that the $g^P$ are about an order of magnitude smaller than the $f^P$.
 Therefore, it is safe to conclude that an interior object will always be attracted to the spherical cavity walls if $\alpha^M \ll \alpha^E$. If one could make a material, for which $\alpha^M$ was large and dominant, the interior object could float inside the spherical shell instead of being attracted to its walls.

There are important differences between Eq.~(\ref{eq:orientation}) and the classic Casimir-Polder result: first, the energy depends in a non-trivial way on $a/R$; second, at any non-zero distance from the center, the interior object experiences a torque; and third, the force between the two bodies depends on the interior object's orientation. 

To explore the orientation dependence of Eq.~(\ref{eq:orientation}) assume, for simplicity, there is a single frame in which both $\alpha^{E}$ and $\alpha^{M}$ are diagonal. In this body-fixed frame, write $\alpha^{0}_{xx}-\alpha^{0}_{yy} = \beta$ and $\alpha^{0}_{zz}-\frac{1}{2}(\alpha^{0}_{xx}+\alpha^{0}_{yy}) = \gamma$ (where we have suppressed the $M/E$ label). The polarizability in the ``lab frame'' is obtained by  $\alpha =\mathcal{R}  \alpha^{0} \mathcal{R}^{-1}$, where $\mathcal{R}$ is a rotation matrix that orients the principal axes of the inner object with respect to the lab frame~\cite{orientation}. This procedure leaves $\text{Tr}\alpha^0$ invariant, and gives for the second line in Eq.~(\ref{eq:orientation}),
\begin{align}
 \sum_{P = M,E}g^{P}(a/R)& \left(\frac{3\beta^{P}}{2} \sin^{2}\theta\,\cos 2\phi +\gamma^P(3\cos^{2}\theta-1)\right) \, ,
\nonumber
\end{align}
where $\phi$ corresponds to the azimuthal rotation of the object about its principal $z$-axis, and $\theta$ is the angle between the object's principal $z$-axis and the ``laboratory'' $z$-axis connecting the center of the sphere to the origin of $\Sigma_{i}$.
 
If $\beta \ne 0$ then the object held at fixed inclination, $\theta$, experiences a torque that causes it to rotate about the body-fixed $z$-axis.  If, however, the object has axial symmetry $(\beta=0$), then the only torque on the object tries to align it either parallel or perpendicular to the displacement axis.

\begin{figure}[t]
\includegraphics[width=8.6cm]{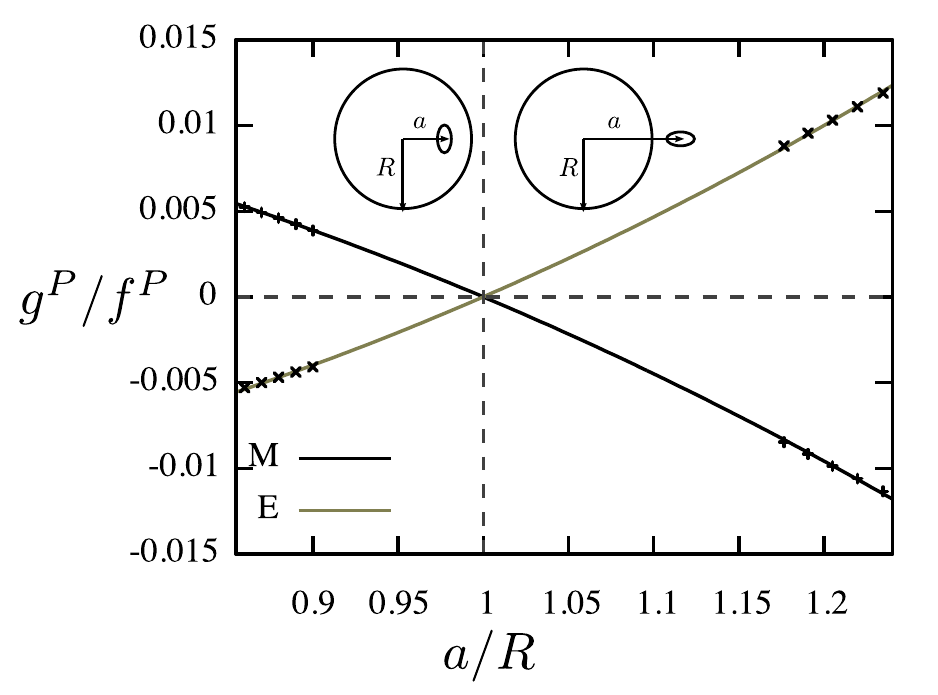} \caption{The ratio $g^P/f^P$, which determines the preferred orientation of the interior object, plotted versus $x=a/R$ showing the change in preferred orientation from interior ($a/R<1$) to exterior ($a/R>1$) (displayed by two small ellipses as described in the text). The data points are numerically computed from Eqs.~\eqref{eq:trace1}, \eqref{eq:trace2}. The solid curves are fits of the form $c_1(1-x)+c_2(1-x)^2$ to these data points. }
\label{deltaoverh}
\end{figure}

A ``cigar shaped'' object ($\gamma>0$) prefers to orient so as to point perpendicular to the $z$ axis, and a  ``pancake'' ($\gamma <0$) tries to align its two large axes perpendicular to the $z$ axis. The small ellipse inside the sphere in Fig.~\ref{deltaoverh} illustrates a side view of both the cigar and the pancake in their preferred orientation. It is interesting to note that $g^E$ and $g^M$ are both positive. So, in contrast to the force, the contributions to the torque from magnetic and electric polarizabilities are in the same direction, if they have the same sign. More complicated behavior is possible if, for example, the electric and magnetic polarizabilities are not diagonal in the same body-fixed coordinate system. Note that our results cannot be compared to the PFA approximation since the the size of the inner object, not the separation of surfaces, $d$, has been assumed to be the smallest scale in the analysis.

An identical analysis can be performed for a polarizable object outside a metallic sphere where $a/R>1$.  The analogous exterior functions $f(a/R)$ and $g(a/R)$ are obtained by exchanging $I_{\nu}$ and $K_{\nu}$ in Eqs.~(\ref{eq:trace1}),~(\ref{eq:trace2}) and~(\ref{eq:gamma}). It turns out that $g(a/R) <0$ for both polarizations. Therefore, the preferred orientation of a polarizable object outside a metallic sphere is opposite of that in the interior case (see the small ellipse outside the large sphere in Fig.~\ref{deltaoverh}). The continuation of the functions $f$ and $g$ from ``interior'' to ``exterior'' is displayed in Fig.~\ref{deltaoverh}, where the transition from one orientation to the other is clear.

In the limit $x=a/R\to 1$, we find that the functions $f^{E/M}(x)$ diverge like $(1-x)^{-4}$, while $g^{E/M}(x)$ diverge only as $(1-x)^{-3}$. These reproduce the well-known Casimir-Polder results for an object facing a plane: the object's energy grows like $1/d^{4}$ and does not depend on its orientation at leading order.

The results, presented up to this point, are accurate in the limit $r/R\to0$, where $r$ is the typical length scale of the interior object. For fixed $r/R$ they become more accurate as the separation from the center, $a/R$, vanishes. For example, for a metallic sphere of radius $r$, at $a/R=0.2$, Eq.~(\ref{eq:orientation}) differs from the exact result by $5 \%$ for $r/R=0.1$ and $20\%$ for $r/R=0.2$.  A more detailed comparison can be found in Ref.~\cite{saad}.

\begin{figure}[t]
\includegraphics[width = 8.6cm]{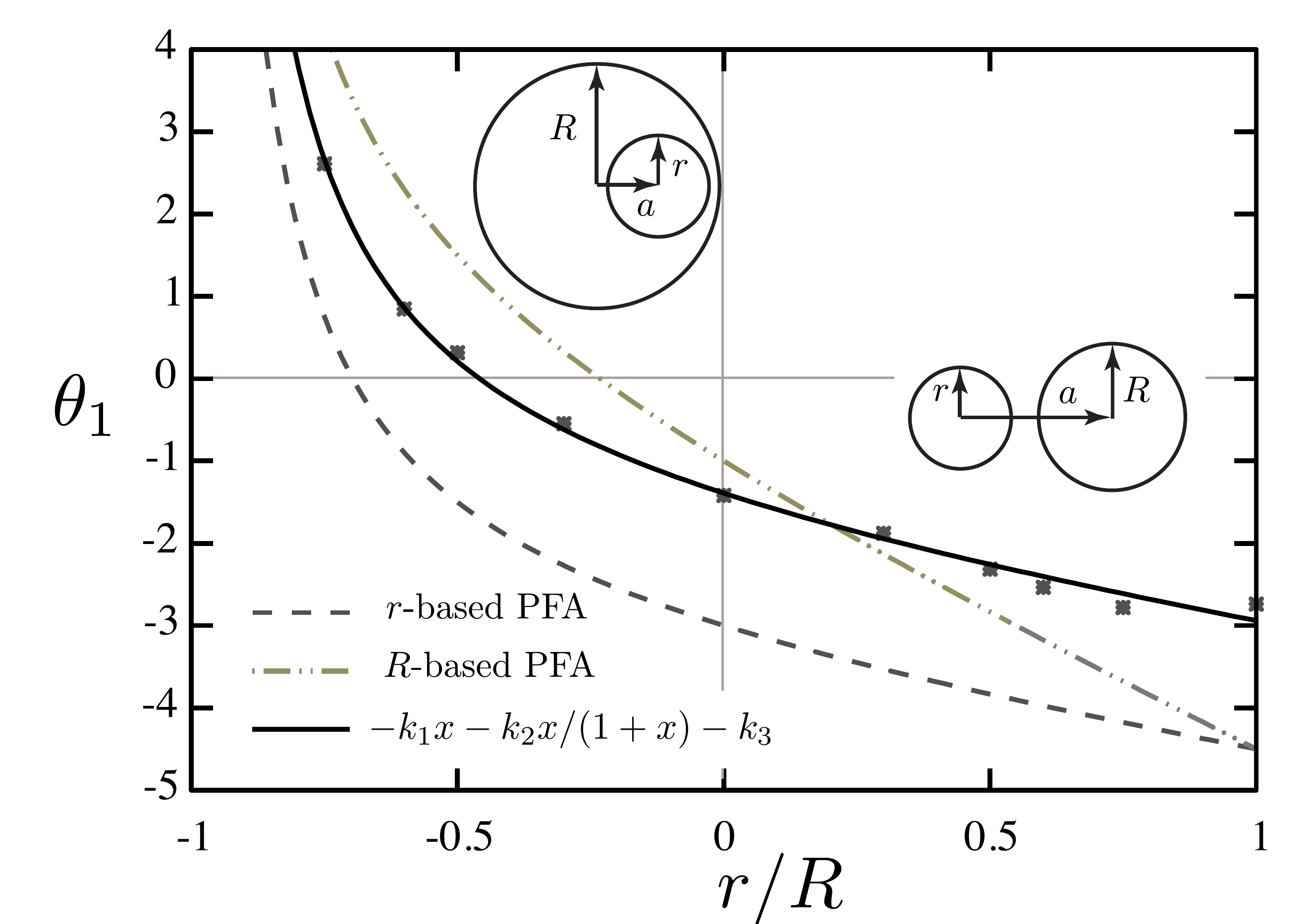} \caption{PFA correction coefficients for spheres. $r/R$ ranges from -1 (interior concentric), to zero (sphere-plane), to $+1$ (exterior, equal radii).  The data points correspond to the exact values of $\theta_1$ calculated numerically, while the solid black curve is a fit (see text). Inset: ``interior'' and ``exterior'' geometrical configurations.}
\label{pfacorrect}
\end{figure}

The opposite extreme from the small-object limit comes when the interior object is nearly touching the cavity wall.  In this limit the Casimir force $F$ between two conducting spheres, which is attractive, is proportional in magnitude to $d^{-3}$, where $d= R-r-a$ is the separation of surfaces. The coefficient of $d^{-3}$ is given by the PFA~\cite{PFA},
\begin{align}
\lim_{d\to 0}d^{3}\, F= - \frac{\pi^3 \hbar c}{360} \frac{rR}{r+R} \, . \label{pfalimit}
\end{align}
This result holds for both the interior and the exterior configuration of two spheres. For fixed $r$ we formally distinguish the cases: $R>0$ for the exterior, $R\to\infty$ for the plate-sphere, and $R<0$ for the interior configuration (see Fig.~\ref{pfacorrect} for reference). All possible configurations are taken into account by considering $-1\le r/R\le 1$.  

Recently, there has been much interest in determining the first correction to the leading PFA result to order $d/r$~\cite{thorstencyl, beyondpfa}. The plate-sphere case and exterior problems of spheres of equal radii have been computed in Refs.~\cite{thorsten} and~\cite{thorstencyl}, respectively. Since most experiments up to now have considered spherical conductors separated by distances much smaller than their radii, the first correction in $d/r$ to the PFA is the geometric correction of greatest practical interest. 

Our numerical studies of the interior configuration for conducting spheres~\cite{saad} enable us to study the limit $d/r \to 0$. This is a difficult limit because no simplifying approximations can be applied to Eq.~(\ref{eq:master}). All powers of $\mathcal{N}$ contribute, and the number of partial waves ($l, l'$) necessary to obtain convergence grows as $d/r\to 0$. Although we know of no derivation of the functional form of the Casimir force beyond the leading term in the PFA, our numerical data are well fit by a power series in $d/r$,
\begin{align*}
F= - \frac{\pi^3 \hbar c}{360 d^3 } \frac{rR}{r+R}\left( 1 + \theta_{1}(r/R)\frac{d}{2r} - \theta_{2}(r/R)\frac{d^{2}}{2r^{2}}+... \right)
\end{align*}
We have used this functional form to extract the coefficient $\theta_{1}(r/R)$. 

It is useful  to have an estimate, however crude, of  $\theta_{1}(r/R)$ over the whole range of $r/R$ with which to compare our results.  Although the PFA is accurate only in the limit $d/r\to 0$, it can be extended in various ways to the whole range of $d$, $r$, and $R$.  The PFA is obtained by considering both surfaces as made up of infinitesimal parallel mirrors. From each point $(\xi_1,\xi_2)$ on the surface of object $O$ one computes the distance $L(\xi_1,\xi_2)$ to the other object's surface along the surface normal $\mathbf{\hat{n}}(\xi_1,\xi_2)$.
By integrating the Casimir energy per unit area for two parallel plates separated by $L(\xi_1,\xi_2)$ over the surface of object $O$ one obtains the ``$O$-based'' PFA energy. Clearly, the result depends on which object one chooses as $O$, but the various results do agree to leading order in $d/r$. We can choose either of the two spheres to arrive at the ``$r$-based PFA'' or the ``$R$-based PFA'', see Fig.~\ref{pfacorrect}. Either one yields a `correction' to the leading order PFA,
\begin{align*} 
\theta^{\text{PFA}}_{1,r}(x) = -\left(\!\!x+\frac{x}{1+x}+3\!\right), \!\!\text{  }\theta^{\text{PFA}}_{1,R} = -\left(\!\!3x +\frac{x}{1+x}+1\!\right),
\end{align*} 
where $x=r/R$. Again, $\theta^\text{PFA}_{1,r}$ and $\theta^\text{PFA}_{1,R}$ are only used for comparison with the actual correction $\theta_1$. Note that the PFA predicts a smooth continuation from the interior to the exterior problem.

In Fig.~\ref{pfacorrect} we plot the values of $\theta_{1}$ extracted from a numerical evaluation of the force from Eq.~\eqref{eq:master} for various values of $r/R<0$, along with the values for $r/R=0$ and $r/R=1$ from Refs.~\cite{thorsten} and \cite{thorstencyl}. We have also repeated the exterior analysis of Ref.~\cite{thorsten} for other values of $r/R>0$. For reference, the two PFA estimates are also shown. 

Eq.~(\ref{eq:master}) is numerically evaluated by truncating the matrix $\mathcal{N}$ at finite multipole order $l$, and extrapolating to obtain the $l \to \infty$ limit. For the data in Fig.~\ref{pfacorrect}, $\mathcal{N}$ was truncated at $l = l' \leq 60$ for the interior and at $l=l' \leq 35$ for the exterior configuration. 
 
The numerical data in Fig.~\ref{pfacorrect} show a smooth transition
from the interior to the exterior configuration. Although the PFA
estimates do not describe the data, the $r$-based PFA has a similar
functional form and divergence as $x\to -1$. Therefore, we fit the
data in Fig.~\ref{pfacorrect} to a function, $\theta_1(x) = -(k_1 x
+k_2x/(1+x) +k_3)$ and find, $k_1=1.05 \pm 0.14, k_2 = 1.08 \pm 0.08,
k_3 = 1.38 \pm 0.06$. This provides a simple form for the leading PFA
correction for metallic spheres, one inside the other and both
outside, which is relevant for many experiments. Notice, however, that
the actual function $\theta_1(x)$ is not known analytically and that
our fit represents a reasonable choice which may not be
unique. Our results show that the correction to the PFA has a
significant dependence on ratio of curvatures of the two surfaces.
The correction is a factor of two larger for two spheres of equal
radii than for the sphere-plane setup; it vanishes near $r/R=-0.5$;
and it becomes large positive as $r/R\to -1$.  These effects should be
taken into account in future experimental searches for PFA
corrections.

This work was supported by the NSF through grant DMR-08-03315 (SJR), DFG through grant EM70/3 (TE),
the U. S. Department of Energy (D.O.E.) under cooperative research
agreement \#DF-FC02-94ER40818 (RLJ), and MIT's undergraduate research
opportunities program (UROP) (SZ). We thank Noah Graham and Mehran
Kardar for many useful conversations regarding this work.

\end{document}